\documentclass[12pt,preprint]{aastex}

\newcommand{\etal}{{\it et al.}}

\begin{document}

\title{Discovery of a low-eccentricity, high-inclination 
Kuiper Belt object at 58~AU}
\author{
R. L. Allen, B. Gladman \\
Department of Physics and Astronomy, 6224 Agricultural Road,\\
University of British Columbia, Vancouver, BC V6T 1Z1, CANADA \\
%\\
J.J. Kavelaars \\
HIA/NRC, Canada\\
%\\
J-M Petit \\
Observatoire de Besan\c{c}on, FRANCE\\
%\\
J. Wm. Parker\\
Southwest Research Institute, USA\\
%\\
P. Nicholson\\
Cornell, USA\\
}
%\affil{.}

%\date{} 		% Activate to display a given date or no date

\begin{abstract}
We report the discovery of the first trans-neptunian object, 
designated 2004~XR190,
with a nearly-cirular orbit beyond the 2:1 mean-motion resonance.
Fitting an orbit to 23 astrometric observations spread out over 12 months 
yields an orbit of $a=57.2\pm0.4$, $e=0.08\pm0.04$, and $i=46.6^{\circ}$.
All viable orbits have perihelia distances $q>49$~AU.
The very high orbital inclination of this extended scattered
disk object might be explained by several models, but its existence
again points to a large as-yet undiscovered population of transneptunian
objects with large orbital perihelia and inclination.
\end{abstract}


%----------------------------------------------------------

\section{Introduction}

Over the last decade, serious observational effort has gone into
detecting Trans-Neptunian Objects (TNOs), but despite increasing resources
dedicated to the problem there is still a steady stream of surprises.
So little light is reflected from distant TNOs
that it is the very inner edge (within 50~AU) of the 
Kuiper Belt region that dominates detections in observational
surveys (those with more than 10 include \citet{JLC96, TJL01, Larsen01, Glad01,
Allen02, Elliot05, Petit05, Allen05}).
The majority of Kuiper belt detections come from these flux-limited surveys
near the ecliptic plane. This results in a bias against objects which are
distant (since reflected flux $\propto d^{-4}$), on highly-inclined orbits (which 
spend very little time in the ecliptic plane), 
or intrinsically rare (like the very largest TNOs). 
These biases are gradually being overcome, and new dynamical classes within
the Kuiper belt are being discovered.

These sub-populations of the Kuiper Belt (see
\citet{Glad05} for a recent review) preserve a record of the
dynamical processes which governed the formation of the giant
planets.
The low-eccentricity `classical belt' appears to decline 
rapidly at the location of the 2:1 mean-motion resonance
\citep{Allen01, TruBro01}
Other mean-motion resonances are occupied both inside
and outside the 2:1, although outside the 2:1 objects are 
only present at high eccentricity (with the possible exception of 2004~XR190).
The scattered disk population of TNOs has perihelia $q<40$~AU
but large semimajor axes $a$; this is a decaying population that
has presumably been flung to large-$a$ via scattering with 
Neptune \citep{dl97}.
Finally, the extended scattered disk (which eventually merges into
the inner Oort Cloud) consists of TNOs on stable orbits, pointing
to some process capable of lowering the orbital eccentricities
of vast numbers of scattered disk objects \citep{Glad02,Sedna04}.

In this Letter we report the discovery of a new extended
scattered-disk object with a nearly-circular orbit beyond
the edge of the classical belt and also with one of the highest
orbital inclinations known.

\section{Observations}

The object reported here was discovered 
in routine data reduction of the Canada-France Ecliptic Plane Survey
(CFEPS) \citep{Allen05}. The internal survey designation
was 0716p004b7 and 
it has received Minor Planet Center designation 2004~XR190 
(in MPEC 2005-X72 \citep{Marsden05}).
In this section, we describe the discovery and tracking observations
up to 2005 December 10.

\subsection{Discovery}

2004~XR190 was discovered in 
images taken as part of the Canada-France Legacy Survey using
data from the Very-Wide 
component (CFLS-VW) on the Canada-France-Hawaii Telescope (CFHT) 
3.5-m and Megacam CCD camera. 
The images were processed by CFHT using their Elixir data processing 
pipeline \citep{Magnier04}, 
and then searched for TNOs with our moving-object 
software \citep{Petit04}. 
2004~XR190 was clearly identified by the software, being about 2 
magnitudes above the limit and uncontaminated by 
background objects. 
A fourth image, taken on the following night, gave a preliminary 
orbit indicating the object's distance was well beyond 50~AU. 
These exposures were taken on 2004, December~11 and 12, using an $r$ filter 
and 90-second exposure times. 
2004~XR190 was measured to have an $r$ magnitude of $21.8\pm0.2$ in
these short exposures. 

Astrometry of 2004~XR190 from the discovery images 
indicated it was at a barycentric distance of $59\pm5$~AU when 
an orbital solution was fit using the \citet{Bernstein00} 
software. This indicates its diameter is between 425--850~km
for a range of albedo of 16--4\%.

Although orbital elements are poorly constrained with only a 
24-hour arc, because the observations were taken at opposition the 
distance estimate from 2-nights of observation was felt to be accurate to 10\%.
Even with the uncertainty, this made 2004~XR190 one of the most distant 
TNOs ever discovered.
Much more uncertain at the time, but even more exciting, was the fact that
its preliminary orbit indicated that $a=59\pm30$~AU, $e=0.02\pm0.5$, and 
$i=45^\circ$ ($\pm24^\circ$).
The unusual nature of this orbit and the large barycentric 
distance led us to follow this object at the next opportunity in 2005. 

\subsection{Tracking}

On 2005 October 3 and 4, a first set of recovery observations was
obtained at the Palomar 5-m using the Large-Format Camera.  
Observations taken one year after a 24-hour arc generally have a very large
ephemeris uncertainty \citep{Allen05}, around $27\arcmin$ in this case. 
However, 2004~XR190 was immediately
visible in the field within $1\arcsec$ of the prediction from the nominal
nearly-circular orbit.  Incorporating the new astrometric measurements
indicated that the distance was indeed $58.6\pm0.3$~AU, although the
best-fit orbit became more eccentric with a higher $a$
($a=63\pm28$~AU, $e=0.3\pm0.7$).  This recovery also confirmed that
2004~XR190 had one of the largest orbital inclinations of any TNO
(then $i=46.1\pm0.4^\circ$); only one other classical KBO and one scattered disk
object have higher inclinations.  This pattern of orbits changing from
near-circular to elliptical often emerges when fitting increasing
arc-lengths with the Bernstein \& Khushalani code, as the orbit
assumptions are changed as the observational arc grows.  However, the
uncertainties remain large and the final orbit remains within these
limits.

Further observations took place in 2005, at the MDM 2.4-m on October 15, 
at the Kitt Peak Mayall 4-m on November 4 and 6,
and again at CFHT on December 1. 
Although these observations only increased the arc-length from 10 to 12 
months, because of the spread  of the observational geometry
relative to opposition, orbital-element uncertainties dropped
very rapidly. 
Including all available observations, we find the best-fit orbit to 
be $a=57.5\pm0.6$~AU, $e=0.11\pm0.04$, and $i=46.641\pm0.005^\circ$,
with node=$252.367^{\circ}$, argument of perihelion $284^\circ\pm6^\circ$, and
time of perihelion passage of JD=$2494000\pm3000$.

\begin{figure}[htbp]
\includegraphics*[width=5in]{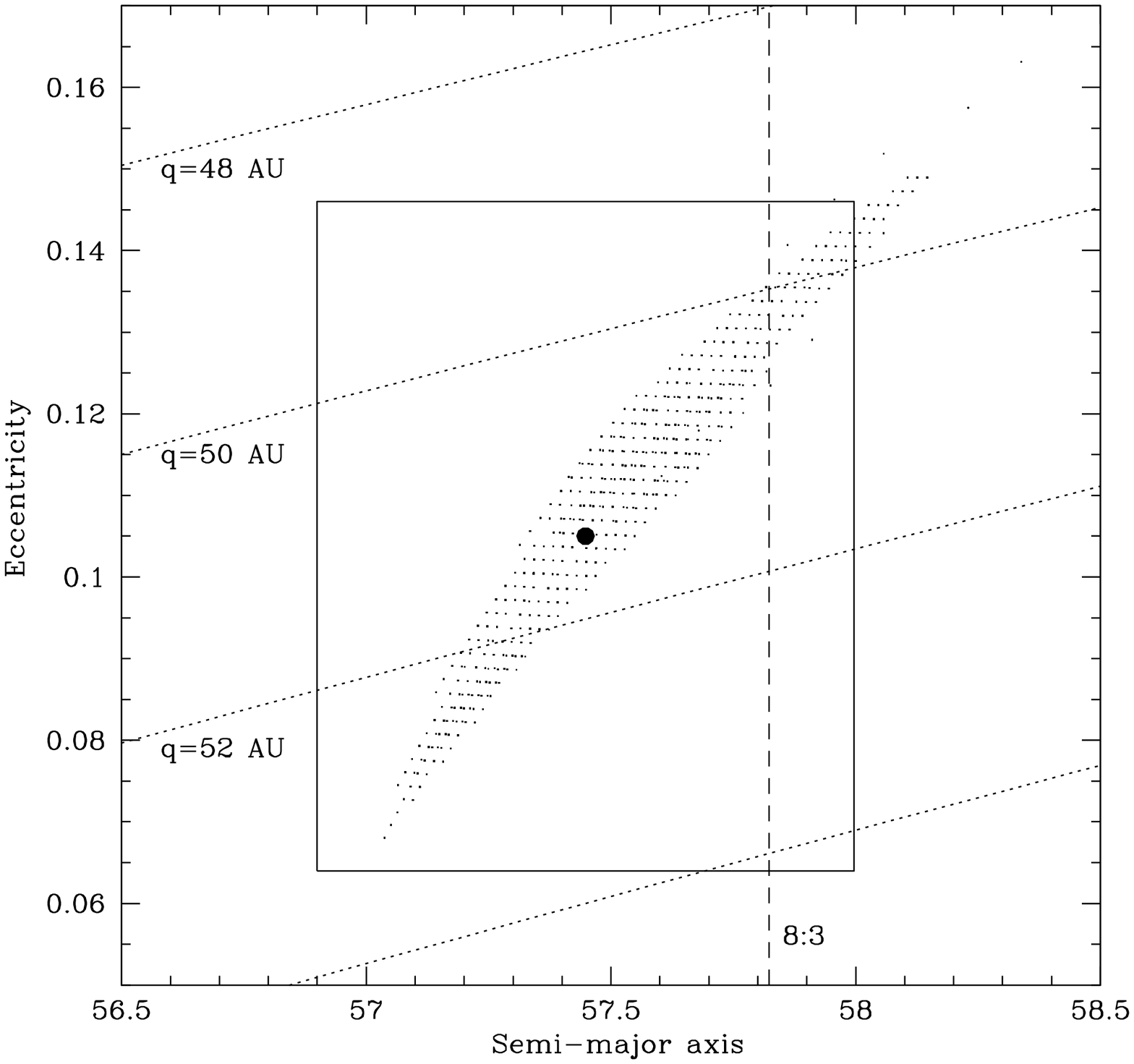}
\caption{The coupled uncertainties in semi-major axis and eccentricity for
2004~XR190. The best-fit orbital elements are shown by a heavy dot.
The small points show orbital solutions consistent with our astrometric data to
a $0.5\arcsec$ mean-residual level. The distribution of these points illustrates 
the correlation between 
$a$ and $e$ and may be considered an uncertainty ellipse for
these parameters. The box shows the 1-sigma uncertainties in $a$ and $e$ returned
by the \citet{Bernstein00} software (stated in the text), but this ignores the $a$/$e$ correlation.
The dashed line indicates the center of the 8:3 mean-motion resonance at 57.8~AU.
Dotted curves are loci of constant perihelion.
2004~XR190's best-fit perihelion distance is at 51.4~AU. 
\label{4b7error} }
\end{figure}

While the large current barycentric distance of $58.43\pm0.03$~AU 
and high inclination are unusual, a 
third, and probably most important, feature of this orbit is its 
large pericentric distance. 
The uncertainties quoted above in $a$ and $e$ are correlated; increases 
in $e$ necessitate an increase in $a$ to continue to fit the 
observations, as illustrated in Figure~\ref{4b7error}.
As such, the limits on $q$ are stricter than the simple one-sigma limits 
above. 
Taking into account this correlation, we find that the lowest possible 
$q$ compatible with the observations, at the one-sigma level, is in 
fact 49.4~AU. 
This places 2004~XR190 in the extended scattered disk, although its 
eccentricity is the lowest of any member of this group. 

%details on other higher i, more distant objects for comparison? other esdos?

% note - add color info here

\section{Discussion}

\begin{figure}[htbp]
\includegraphics*[width=5.5in]{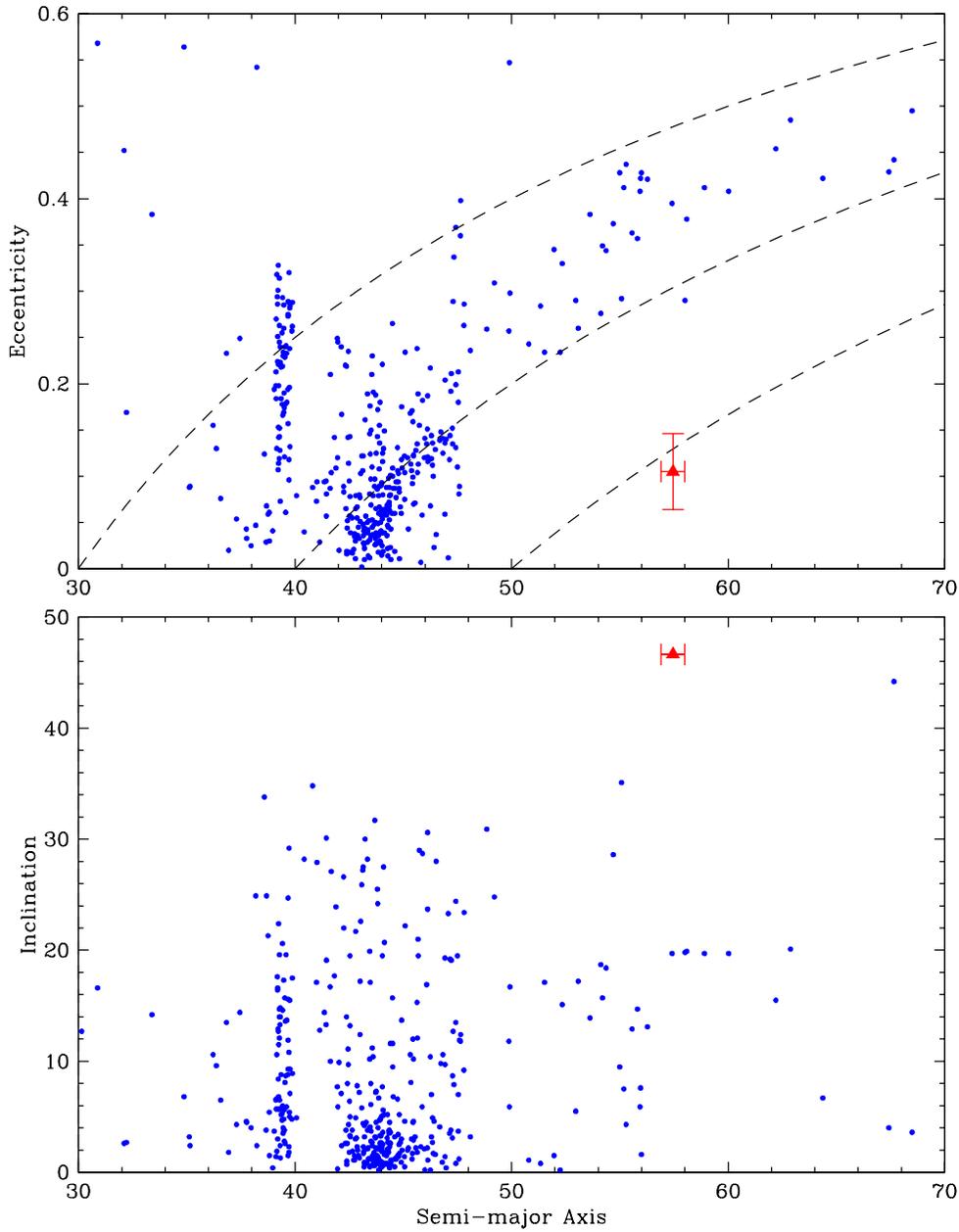}
\caption{Points show well-determined orbits for Kuiper belt objects in
semi-major axis vs. eccentricity and semi-major axis
vs. inclination space. 2004~XR190's current best-fit orbit is indicated 
by a triangle with error bars
for $a$ and $e$/$i$. Objects from the Minor Planet Center database
with an observational arc of over 2 oppositions are indicated by
dots. Dashed curves on the semi-major axis vs. eccentricity plot
indicate perihelion distances of 30, 40 and 50~AU. 
\label{aeifig} }
\end{figure}

2004~XR190 is unlike any other member of the Kuiper belt, due to its 
high pericenter and highly-inclined orbit, as illustrated in 
Figure~\ref{aeifig}.
With a fairly circular orbit beyond 50~AU, it would be tempting to think 
of this object as the first member of a ``cold distant belt'' 
(see \citet{Stern97, Hahn99}). 
However, its high inclination suggests that it has experienced a strong 
dynamical perturbation in its history. 

We discovered 2004~XR190 only $\sim1^\circ$ away from the ecliptic plane. 
With $i=47^\circ$, it spends only a tiny fraction ($\sim2\%$) of its orbit 
within this limit.
Most TNO surveys do not extend further than a few degrees from the ecliptic 
plane, and so have poor sensitivity to TNOs on such high-$i$ orbits.
The on-going Caltech Survey\citep{TrujilloBrown03} has covered a major
fraction of the sky within 10 degrees of the ecliptic, and is
increasing coverage further away. The limit of this survey in in the
range $m_R\sim20$--$21$, such that objects must be 
Pluto-scale or larger to be detected beyond a distance of 60~AU.  With these
selection effects, we cannot rule out a large population of high-$i$
objects like 2004~XR190.  However, a population with similar $a$ and
$e$ but low-$i$ should have been undetected in prior surveys.  Indeed,
\citet{Allen02} and \citet{TruBro01} sets strong limits on such a
distant `cold belt'.  Therefore, we conclude that 2004~XR190 does not
represent the high-inclination end of a dynamically cold population beyond
50~AU. More likely is that this discovery is a member of an as-yet
poorly characterized very high-$i$ group.  Because of the presence of
other very highly-inclined TNOs in the classical belt and the
scattered disk, it is unclear if the highest-inclination population is
especially concentrated in the extended scattered disk or if it
extends throughout the Kuiper belt.  

%Likewise, there is a bias towards finding objects with smaller perihelia before similar objects are detected at greater distances. 

Placing a TNO onto a nearly-circular orbit near 60~AU with a high 
inclination, while simulataneously leaving intact the inner Kuiper belt
and depopulating the low inclination orbits beyond the 2:1 resonance,
%without simulataneously having a large number of similar TNOs on 
%low-$i$ orbits 
is a challenge for theories seeking to create the 
extended scattered disk.
These theories include close stellar passages, rogue planets/planetary 
embryos in the early Kuiper belt, and resonance interaction with a 
migrating Neptune. 

In stellar passage models, a star has a close encounter with the primordial
Kuiper belt or scattered disk. 
The  end result of these encounters is a Kuiper belt which transitions from 
a slightly-perturbed  to a greatly-perturbed state beyond some critical 
distance\citep{Ida00, Kenyon04, Morbi04, Koba05}.  Generally, objects produced 
with high inclination in these models also have high eccentricities, 
which make it difficult to produce 2004~XR190.
Typically the stellar passage scenarios leave behind an extended
scattered disk in which the mean inclination of the Extended Scattered
Disk objects (ESDOs) rises
as one moves to larger semimajor axes. With the addition of 2004~XR190 
to the suite of ESDOs 1995~TL8, 2000~YW134, 2000~CR105, and 2003~VB12/Sedna 
\citep{Glad02, MorbiEmel04}, the current trend
appears to be the opposite. While the lower-$a$ ESDOs should indeed be detected
in greater numbers first (due to distance/flux detection biases), the lack 
of high-$i$ ESDOs with $a>100$~AU might be viewed as a problem for 
stellar passage models.

Recent simulations by \citet{GladChan06} show that a 1--2 Earth mass rogue
planet living temporarily in the scattered disk can effectively create
high-$q$ ESDOs.
Production of objects with orbital inclinations above 40 degrees
is possible but not efficient, and the high-$i$ objects that are created
tend to be at the lowest semimajor axes ($a<100$~AU).
If TNOs with orbital inclinations above $30^{\circ}$ are discovered
beyond $a=100$~AU then a stellar passage scenario should indeed be
favoured. If this method did produce 2004~XR190, $a \sim 100$~AU, $e \sim 0.5$ TNOs 
should soon be discovered with inclinations between 10--40$^\circ$.  

An intriguing explanation is that 2004~XR190 evolved to its current orbit after 
being trapped in a mean-motion resonance with Neptune while Neptune migrated 
outwards. 
In this model, TNOs are trapped into resonance 
and evolve onto higher semi-major axis orbits as Neptune migrates 
outward, increasing their eccentricities in the process \citep{Hahn05}. 
This process is inefficient at increasing orbital inclinations to large
values if the trapped objects begin on circular orbits. However,. 
a similar effect can occur if already-eccentric TNOs become trapped in the resonance
during Neptune's migration. This can result in inclination pumping, primarily due
to the Kozai resonance. 
Although weak in the Kuiper belt outside of 
mean-motion resonances, inside or near the edges of these resonances the Kozai effect
is capable of transferring the orbital
eccentricity into an elevated inclination \citep{Gomes03, Gomes05}.
If the object then drops out of the resonance, it will be left on 
a high-$i$, low-$e$ orbit relatively stable against the gravitational 
perturbations of Neptune and the other giant planets.
This could explain 2004~XR190's orbit, without requiring low-$i$ 
objects of similar semi-major axis. 
\citet{Gomes05} show an example of this process using the 5:2 mean-motion
resonance, but this resonance is interior to the best-estimate semimajor
axis of 2004~XR190, so use of the 5:2 resonance to produce the orbit would
require a final inward migration of Neptune.
This is not impossible, as \citet{Gomes04}
show that both inward and outward Neptune migration can occur.
2004~XR190's best-fit orbit is in fact closer to the 8:3
than the 5:2 mean motion resonance. Unfortunately, the ability of the 8:3 resonance to 
participate in this mechanism has not been demonstrated. No published numerical
models appear to exhibit trapping and strong inclination pumping in the 
8:3 resonance. 

\begin{figure}[htbp]
\includegraphics*[width=5.5in]{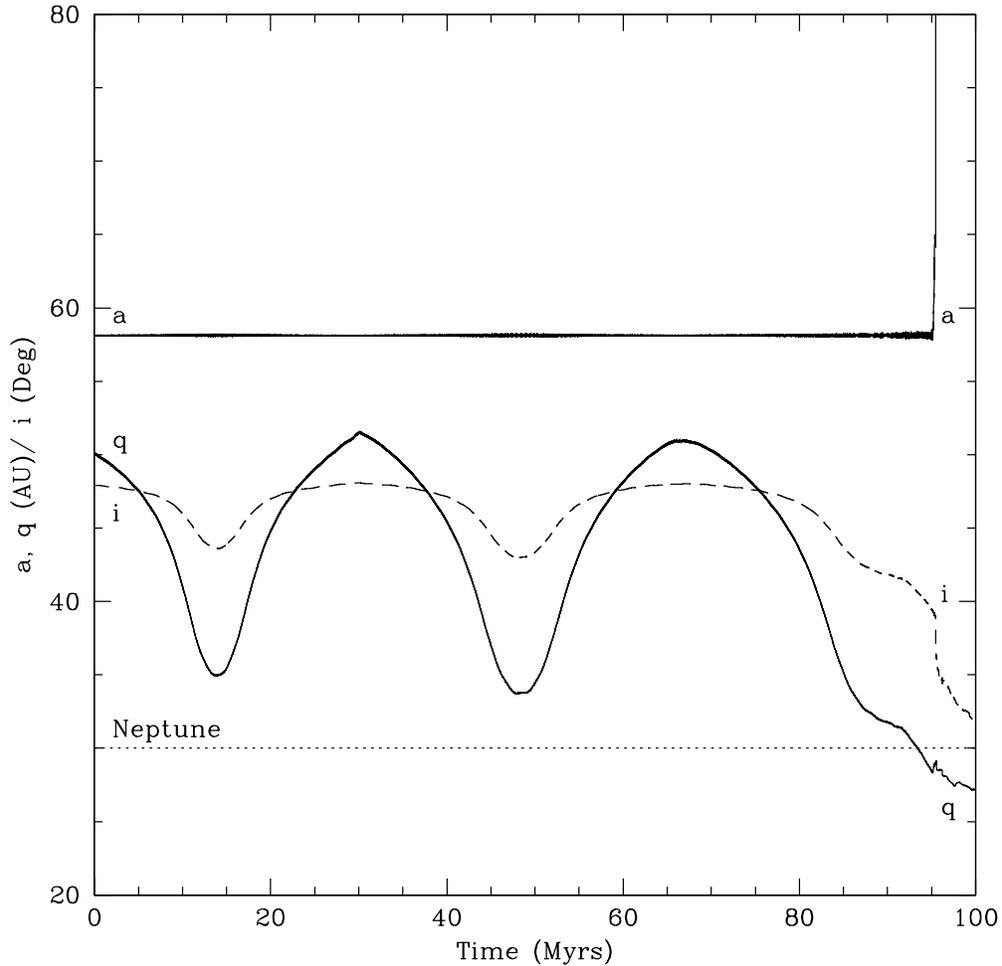}
\caption{
Evolution of a clone interacting with the 8:3 resonance. The particle's intial orbital elements 
are within 2004~XR190's orbital uncertainties, but place it near the resonance..
This particle undergoes Kozai pumping of its eccentricity such that its perihelion
distance periodically drops below 35~AU and during the third eccentricity minimum interacts
strongly with Neptune and is removed from near the resonance.
The solid
curves indicate semimajor axis (a) and perihelion (q).The dashed curve indicates the inclination (i)
history, which is in phase with the q history due to the Kozai effect.
Neptune's semimajor axis throughout the simulation is indicated by the dotted curve.
\label{resonantfig}
}
\end{figure}

We have conducted preliminary orbital integrations of 
a suite of particles consistent with the 1-sigma uncertainties
of 2004~XR190's orbital elements. We integrated 100 clones for $10^7$ years. 
The vast majority of these clones show small ($\sim0.01$) variations in eccentricity
over this time period. 
However, given the current uncertainties in the orbit, we find that there is a 
roughly 5\% chance (see Fig.~\ref{4b7error}) that XR190 has
orbital elements that would allow it to be strongly influenced by the 8:3 resonance.
The resonant particles (as diagnosed by libration
of the angle
$\phi = 8\lambda - 3\lambda_N - 5\tilde{\omega}$
during the intgrations)
show an extremely strong Kozai response which brings the
perihelia down into the scattered disk region ($q<38$~AU)
within $10^7$ years. We extended the integration of the resonant
clones to $10^8$ years. The behavior of one of these
resonant particles is illustrated in Figure~\ref{resonantfig}; in
this case the particle actually crosses Neptune's orbit after 80Myrs.
We have confirmed that $e$, $i$, and $\omega$ show the
correct coupled Kozai behavior.
This raises the possibility, if future observations result
in a resonant orbit, that XR190 is only temporarily resident
in the low-eccentricity domain.
In this scenario, 2004~XR190 could have recently (as little as
a few tens of millions of years ago) been a scattered disk
object, which is simply undergoing a low-eccentricity episode
by virtue of having been fortuitously near the resonance boundary
after a Neptune-scattering event.
The dynamics of this near-resonance should be further explored if
observations in early 2006 confirm that $a$ and $e$ are both
at the high end of the currently-allowed range.
% the semimajor axis is indeed in the 8:3.
Even if future observations show 2004~XR190 has $a$ just below that
of the resonance (as the current best-fit orbit indicates), the strong 
resonant response seen in our integrations does suggest the possiblity 
that a 'resonant dropoff' mechanism could have delivered it to its
current location.

\section{Conclusion}

We have presented the discovery of an unusual TNO, 2004~XR190. 
With a perihelion at $\sim50$~AU, and a low-eccentricity orbit, it is 
the closest to a `distant cold Kuiper belt' object which has 
been detected so far. 
However, with an inclination of $47^\circ$, it has clearly been 
dynamically perturbed at some point in its lifetime. 

A plausible explanation of the origin of 2004~XR190's high inclination 
and low eccentricity is the action of the Kozai effect during a
past residence inside the 5:2 or 8:3 mean-motion resonances 
of Neptune. 
If Neptune migrated outwards, dropping the TNO out of resonance, this 
could aid in freezing the $e$/$i$ combination observed today. 
The modification of its orbit could also be produced by now-absent
bodies (rogue planets or passing stars), but producing all of the
features present in the transneptunian region is problematic for
all of the above models.

\section{Acknowledgements}
This paper would not be possible without observations obtained at a number
of telescopes.

Based on observations obtained with MegaPrime/Megacam, a joint project
of CFHT and CAE/DAPNIA, at the Canada-France-Hawaii Telescope (CFHT)
which is operated by the National Research Council (NRC) of Canada,
the Institut National des Science de l'Univers of the Centre National
de la Recherche Scientifique (CNRS) of France, and the University of
Hawaii. This work is based in part on data products produced at
TERAPIX and the Canadian Astronomy Data Centre as part of the
Canada-France-Hawaii Telescope Legacy Survey, a collaborative project
of NRC and CNRS. Also based partly on observations obtained at the
Hale Telescope, Palomar Observatory, as part of a collaborative
agreement between the California Institute of Technology, its
divisions Caltech Optical Observatories and the Jet Propulsion
Laboratory (operated for NASA), and Cornell University. 
Additional observations from Kitt Peak National
Observatory, National Optical Astronomy Observatory, which is operated
by the Association of Universities for Research in Astronomy,
Inc. (AURA) under cooperative agreement with the National Science
Foundation.

We would also like to thank Dr. Arlin Crotts and Patrick Cseresnjes, Columbia University,
for their help in obtaining images of 2004~XR190. Lynne Allen and Brett Gladman
acknowledge support from NSERC and CFI. Joel Parker acknowledges 
support by NASA Planetary Astronomy Program grant NNG04GI29G.

\end{document}